\documentclass[12pt]{article}

\usepackage[utf8]{inputenc}

\usepackage{indentfirst}
\usepackage{amsmath}
\usepackage{amssymb}
\usepackage{amsfonts}
\usepackage{amscd}
\usepackage{amsbsy}
\usepackage{amsthm}
\usepackage{latexsym}
\usepackage{bbm}
\usepackage{dsfont}
\usepackage{graphicx,color} 
\usepackage[dvipsnames]{xcolor}
\usepackage[colorlinks]{hyperref}
\hypersetup{linkcolor=blue,citecolor=blue,urlcolor=blue}
\usepackage[utf8]{inputenc}
\usepackage{tikz}
\usepackage[compat=1.1.0]{tikz-feynman}
\usepackage{amsmath}
\def\nn{\nonumber}       
\def\beq{\begin{eqnarray}}
\def\eeq{\end{eqnarray}}
\def\ln{\,\mbox{ln}\,}

\def\det{\,\mbox{det}\,}
\def\tr{\,\mbox{tr}\,}
\def\str{\,\mbox{str}\,}
\def\diag{\,\mbox{diag}\,}
\def\Tr{\,\mbox{Tr}\,}

\renewcommand{\Re}{\,\mbox{Re}\,}

\renewcommand{\Im}{\,\mbox{Im}\,}

\def\al{\alpha}
\def\be{\beta}

\def\vp{\varepsilon}
\def\ep{\epsilon}
\def\ze{\zeta}

\def\la{\lambda}

\def\pa{\partial}

\def\ph{\varphi}

\def\Ga{\Gamma}
\def\De{\Delta}
\def\La{\Lambda}

\usepackage[symbol]{footmisc} 
\usepackage{float} 
\usepackage{cite}  
\usepackage{titlesec}

\titleformat*{\section}{\large\bfseries}
\titleformat*{\subsection}{\normalsize\bfseries}
\usepackage{orcidlink}
\begin{document}

\begin{center}
\renewcommand*{\thefootnote}{\fnsymbol{footnote}} 
{\Large \bf
Explicit Symmetry Breaking in Non-Abelian \\
Scalar Theories and One-Loop Renormalization}
\vskip 6mm

{\bf Lucas Ducastelo de C. C. Lima}\,\orcidlink{0009-0009-0344-2035}%
\footnote{E-mail address: \ ldcc.lima@unesp.br}
\vskip 6mm
Universidade Estadual Paulista (Unesp), Faculdade de Engenharia e Ciências de Guaratinguetá, Av. Dr. Ariberto Pereira da Cunha, 333,
\\
12516-410, Guaratinguet\'a, SP, Brazil.
\end{center}
\vskip 2mm
\vskip 2mm


\begin{abstract}
\noindent
 We investigate the one-loop renormalization of a Yang--Mills--Faddeev--Popov theory deformed by a complex scalar doublet with explicitly broken $SU(2)$ gauge symmetry. At the classical level, we map the complex doublet into a real $SO(4)$ multiplet and discuss its generalization to $SU(N)$ through $SO(2N)$. The symmetry breaking is introduced by splitting the scalar masses and quartic couplings. Using the background field method and the Schwinger--DeWitt technique, we compute the one-loop divergences and establish renormalizability at this order. We derive the beta functions, recovering the known gauge-invariant limit, and analyze the RG flow. Although the Yang--Mills coupling remains asymptotically free, the rescaled quartic couplings have no real fixed points, and the one-loop flow develops a divergence at a finite energy scale, even in the symmetric limit. The asymmetry between the rescaled scalar self-couplings increases toward the UV and decreases toward the IR, within the region of nonnegative scalar self-couplings.
\noindent

\vskip 3mm

\vskip 3mm

\noindent

\end{abstract}
\setcounter{footnote}{0} 
\renewcommand*{\thefootnote}{\arabic{footnote}} 

\section{Introduction}
\label{sec1}

Non-Abelian gauge theories, or Yang--Mills (YM) theories, are part of the Standard Model of particle physics and Grand Unified Theories (GUTs). One of their properties is asymptotic freedom, independently discovered by Gross and Wilczek \cite{Gross1973} and by Politzer \cite{Politzer1973}. In the pure Yang--Mills sector, this property causes the coupling to decrease at high energies, ensuring the ultraviolet (UV) stability of this sector.

However, the inclusion of fundamental scalar fields, required for spontaneous symmetry breaking and mass generation through the Higgs mechanism \cite{Higgs:1964ia,Englert:1964et}, can compromise this stability \cite{ChengEichtenLi1974}. Scalar self-interactions can grow with the energy scale, destroying asymptotic freedom and driving the theory toward a Landau pole \cite{ZinnJustin1994}. This behavior constrains the Higgs boson mass and complicates the construction of inflationary models based on the scalar sector of the Standard Model \cite{Bezrukov2008, Barvinsky2008, Bezrukov2009a, Barvinsky2009}.

The quantization of YM theories requires an appropriate treatment of gauge redundancy \cite{Faddeev1980, Itzykson1980}. The initial difficulties concerning the unitarity of loop diagrams were pointed out by Feynman \cite{Feynman1963}, who proposed introducing fictitious fields, known as ghosts, to cancel unphysical degrees of freedom. Shortly afterward, Faddeev and Popov \cite{Faddeev1967} and, independently, DeWitt \cite{DeWitt1967} established a quantization framework based on functional integration over gauge orbits and derived the rules for these ghost interactions. In this context, the background field method (BFM) allows one to calculate effective actions by splitting the fields into classical backgrounds and quantum fluctuations \cite{DeWitt1967, Abbott1981}. In gauge-invariant theories, this method preserves invariance under background field transformations \cite{Barvinsky2018, Grassi2003, Arefeva1975}. In the deformed model considered here, we use the BFM to organize the expansion in the quantum fields.

Building on a recent study of an explicit gauge symmetry-breaking mechanism in scalar quantum electrodynamics (sQED) \cite{Lima2025}, we investigate its extension to the non-Abelian regime. Specifically, we explore the quantum effects and the renormalization group (RG) flow of an $SU(2)$ Yang--Mills--Faddeev--Popov theory coupled to a complex scalar doublet. We first show that the complex doublet can be mapped into a real $SO(4)$ representation, providing a formal generalization to $SO(2N)$ multiplets. Within this setup, following the approach proposed in our previous work \cite{Lima2025}, we introduce explicit gauge symmetry breaking exclusively in the scalar sector by splitting the mass parameters and quartic self-coupling constants.

At the quantum level, we use the background field method combined with the Schwinger--DeWitt technique \cite{DeWitt1965, DeWitt2003} to obtain the one-loop divergences of a Yang--Mills--Faddeev--Popov theory deformed by the addition of a complex scalar sector with explicit symmetry breaking. We demonstrate the one-loop renormalizability of this deformed model and derive the corresponding beta functions. Finally, we analyze the fixed points of the RG flow and the high-energy behavior of the couplings. In the model studied here, the UV flow amplifies the asymmetry between the scalar couplings and leads to Landau poles.

\section{Classical action and gauge invariance}
\label{sec2}
Our starting point will be the non-Abelian gauge theory for a complex scalar field in the fundamental representation of $SU(2)$,
\beq
S_c(\phi) \,\,=\,\,
\int d^4x \Big\{(D^\mu\phi)^\dagger \,(D_\mu\phi)
- m^2 \phi^\dagger\phi
- \la\,(\phi^\dagger\phi)^2 \Big\},
\label{acaocomplex}
\eeq
where $m$ is the mass of the scalar field, $\la$ is the self-interaction constant, and $\phi$ is a doublet of complex scalar fields. Furthermore, the covariant derivatives are 
\beq
&&
D_\mu\phi = \pa_\mu \phi - ig \frac{t^a}{2}A_\mu^a\phi
\qquad
\mbox{and}
\qquad
(D_\mu\phi)^\dagger = \pa_\mu \phi^\dagger + ig \phi^\dagger A_\mu^a\frac{t^a}{2}.
\eeq
Here, $t^a$ are the Pauli matrices, with $a = 1, 2, 3$.
The transformations of the fields that ensure the gauge invariance of Eq.~\eqref{acaocomplex} are
\beq
{A'}_\mu \;=\; \omega A_\mu \omega^{-1}+ \frac{i}{g}\omega \pa_\mu \omega^{-1}
\quad
\mbox{and}
\quad
\phi' \;=\; \omega \phi ,
\quad
\mbox{with}
\quad
\omega = \omega(x) \in SU(2). 
\eeq
We know that $\omega$ can be written in terms of three independent parameters:
\begin{align}\omega = 
    \begin{pmatrix}
       \al &\be \\
       -\be^* & \al^*
    \end{pmatrix},
\quad\text{with } \al,\be \in \mathbb{C} \quad\text{and}\quad
|\al|^2 + |\be|^2 = 1.
\end{align}
We can reparameterize $\omega$ in terms of $x, y \text{ and } z$,
\begin{align}
    \al = e^{i\frac{x+y}{2}}\cos \frac{z}{2} \quad \text{and}\quad \be = e^{i\frac{x-y}{2}}\sin \frac{z}{2}\,.
\end{align}
Thus, the transformation for each complex field of the doublet is given by
\begin{align}
    \begin{pmatrix}
        \phi_1'\\
        \phi_2'
    \end{pmatrix}=
    \begin{pmatrix}
      e^{i\frac{x+y}{2}}\cos \frac{z}{2} & e^{i\frac{x-y}{2}}\sin \frac{z}{2}\\
      -e^{-i\frac{x-y}{2}}\sin \frac{z}{2}&e^{-i\frac{x+y}{2}}\cos \frac{z}{2}
    \end{pmatrix}
    \begin{pmatrix}
        \phi_1\\
        \phi_2
    \end{pmatrix}.
\end{align}
Let us express each complex field of the doublet in terms of two real scalars,  
\beq
\phi_j \;=\; \frac{1}{\sqrt{2}}\left(\ph_j \, + \, i \chi_j\right).
\label{c2r}
\eeq
The transformations of the four real scalars that ensure the gauge symmetry of the action can be written as: 
\begin{align}\label{transf}
    \begin{pmatrix}
        \ph_1'\\
        \chi_1'\\
        \ph_2'\\
        \chi_2'
    \end{pmatrix}=
    Q
    \begin{pmatrix}
        \ph_1\\
        \chi_1\\
        \ph_2\\
        \chi_2
    \end{pmatrix},
\end{align}
where
\begin{align}
    Q = \begin{pmatrix}
      A\cos \frac{z}{2} & B\sin \frac{z}{2}\\
      -B^{-1}\sin \frac{z}{2}&A^{-1}\cos \frac{z}{2}
    \end{pmatrix},
    \end{align}
with
    \begin{align}A =
    \begin{pmatrix}
      \cos \frac{x+y}{2} & -\sin\frac{x+y}{2}\\
      \sin\frac{x+y}{2} & \cos \frac{x+y}{2}
    \end{pmatrix}
    \quad \text{ and }\quad B=
    \begin{pmatrix}
      \cos \frac{x-y}{2} & -\sin\frac{x-y}{2}\\
      \sin\frac{x-y}{2} & \cos \frac{x-y}{2}
    \end{pmatrix}, 
     \text{ } A,B \in SO(2). 
\end{align}
Using the Schur complement formula, since rotation matrices are invertible, we can calculate the determinant of the transformation matrix of real scalar fields, Eq. \eqref{transf}, by taking the determinant by blocks. Thus,
\begin{align}
    \det Q = \det\left(A\cos\frac{z}{2}\right)\det\left(A^{-1}\cos\frac{z}{2} + B^{-1}\sin \frac{z}{2}A^{-1}\sec\frac{z}{2}B\sin \frac{z}{2}\right) = 1.
\end{align}
Here, we use the fact that matrices in $SO(2)$ commute. In addition, taking the transpose of $Q$ and remembering that for elements of the group $SO(2)$, $A^T = A^{-1}$, we have that 
\begin{align}
    &Q^T = \begin{pmatrix}
      A^T\cos \frac{z}{2} & -B\sin \frac{z}{2}\\
      B^{-1}\sin \frac{z}{2}&A\cos \frac{z}{2}
    \end{pmatrix}\Rightarrow\\\nn
    &\Rightarrow Q^TQ = \begin{pmatrix}
      A^T\cos \frac{z}{2} & -B\sin \frac{z}{2}\\
      B^{-1}\sin \frac{z}{2}&A\cos \frac{z}{2}
    \end{pmatrix}
    \begin{pmatrix}
      A\cos \frac{z}{2} & B\sin \frac{z}{2}\\
      -B^{-1}\sin \frac{z}{2}&A^T\cos \frac{z}{2}
    \end{pmatrix} = 1.
\end{align}
Therefore, we conclude that the four real fields transform under transformations of the group $SO(4)$.
\section{Generalization for SU(N)}
\label{sec3}
Before beginning the analysis of quantum theory, let us extend the model to the group $SU(N)$. In this case, we will work with the action \eqref{acaocomplex}, now in the context of $SU(N)$. For this group, we have a multiplet of dimension $N$, whose transformation is given by
\begin{align}\label{transformacaosun}
    \phi' = \omega \phi, \quad \omega = \omega(x) \in SU(N).
\end{align}
In order to simplify operations, we can separate the $\omega$ matrix as
\begin{align}\label{transformacaosunreais}
    \omega = \Re\omega + i\,\Im\omega,
\end{align}
and seek the relationship between these two parts, given that $\omega^\dagger \omega = 1$. Thus, from 
\eqref{transformacaosunreais}, we obtain:
\begin{align}
    \omega^\dagger\omega &=\left[\Re\omega^T - i\,\Im\omega^T\right]\left[\Re\omega + i\,\Im\omega\right] = 1 \quad \Rightarrow \nn \\
    &\Rightarrow 
\begin{cases}
 \Re\omega^T\Re\omega + \Im\omega^T\Im\omega = 1 \\
 \Re\omega^T\Im\omega = \Im\omega^T\Re\omega
\end{cases}\label{imre}.
\end{align}
As in the case of $SU(2)$, we rewrite the complex fields of the multiplet as two real scalar fields,
\beq
\phi_j \;=\; \frac{1}{\sqrt{2}}\left(\ph_j \, + \, i \chi_j\right).
\label{c2r1}
\eeq
We are now in a position to determine the transformations for the real fields, i.e.,
\begin{align}
    \ph' \, + \, i \chi' = \omega\,(\ph \, + \, i \chi) \quad
    \Rightarrow \quad\begin{pmatrix}
        \ph'\\
        \chi'
    \end{pmatrix}
    =
    \tilde{Q}
    \begin{pmatrix}
        \ph\\
        \chi
    \end{pmatrix},
\end{align}
where $\tilde{Q}$ is the $2N \times 2N$ transformation matrix and the real field vector $\begin{pmatrix} \varphi \\ \chi \end{pmatrix}$ is a $2N \times 1$ column matrix. The transformation matrix $\tilde{Q}$ is given by
\begin{align}
\tilde{Q} \,=\,\begin{pmatrix}
      \Re\omega & -\Im\omega\\
      \Im\omega & \Re\omega
    \end{pmatrix}\,.
\end{align}
The transformation matrix $\tilde{Q}$ has determinant one. 
To show this, we apply the Schur complement formula to the block decomposition of $\tilde{Q}$ under the assumption that $\Re\omega$ is invertible, and use the relations in \eqref{imre}. This gives
\begin{align}
    \det \tilde{Q}
    &= \det(\Re\omega)\,
       \det\!\left[\Re\omega + \Im\omega\,(\Re\omega)^{-1}\Im\omega\right] \nonumber\\
    &= \det\!\left[\Re\omega^{T}\Re\omega 
       + \Re\omega^{T}\Im\omega\,(\Re\omega)^{-1}\Im\omega\right] \nonumber\\
    &= \det\!\left[\Re\omega^{T}\Re\omega 
       + \Im\omega^{T}\Im\omega\right] 
       = 1 .
\end{align}
Moreover, the transpose of the matrix $\tilde{Q}$ is given by 
\begin{align}
    \tilde{Q}^T = \begin{pmatrix}
      \Re\omega & -\Im\omega\\
      \Im\omega&\Re\omega
    \end{pmatrix}^T
    =
    \begin{pmatrix}
      \Re\omega^T & \Im\omega^T\\
      -\Im\omega^T & \Re\omega^T
    \end{pmatrix}
\end{align}
and
\begin{align}
  \tilde{Q}^T \tilde{Q} \,=\, \begin{pmatrix}
      \Re\omega^T & \Im\omega^T\\
      -\Im\omega^T & \Re\omega^T
    \end{pmatrix}
    \begin{pmatrix}
      \Re\omega & -\Im\omega\\
      \Im\omega & \Re\omega
    \end{pmatrix}
    =
    \begin{pmatrix}
      \mathbbm{1}_{N\times N} & 0\\
      0 & \mathbbm{1}_{N\times N} 
    \end{pmatrix}
    = \mathbbm{1}_{2N\times 2N}.
\end{align}

Therefore, the transformation matrix $\tilde{Q}$ is orthogonal with determinant one, and hence belongs to the group $SO(2N)$.

Consequently, the generalization follows: complex scalar fields transforming under $SU(N)$ can be mapped into real scalar fields transforming under $SO(2N)$.

Although we have demonstrated that the maximal symmetry of the pure scalar sector can be interpreted geometrically in the real space $SO(2N)$, in the subsequent sections we will maintain the complex fields formalism and the fundamental representation of $SU(N)$. This methodological choice is more convenient for the implementation of the Background Field Method and for the computation of quantum divergences, since it maintains the compact structure of the covariant derivative and simplifies the interactions with the gauge bosons. In this context, the symmetry breaking that we will introduce next can be understood not only as the breaking of the $SU(2)$ gauge, but also as the explicit breaking of the global $SO(4)$ symmetry underlying the scalar sector.

\section{Gauge symmetry breaking}
We can break the gauge symmetry of the action \eqref{acaocomplex} by introducing different and independent masses and coupling constants in the scalar self-interaction sector. Thus, the symmetry-broken action takes the form
\begin{align}
 S_{\text{bs}}(\phi) \,\,&=\,\,
\int d^4x \Big\{(D^\mu\phi_1)^* \,(D_\mu\phi_1)
+(D^\mu\phi_2)^* \,(D_\mu\phi_2)- m_1^2\phi^*_1\phi_1 - m_2^2\phi^*_2\phi_2\nn\\
&\quad- \la_1(\phi^*_1  \phi_1)^2- \la_2(\phi^*_2  \phi_2)^2 - \la_{12}(\phi^*_1  \phi_1)(\phi^*_2  \phi_2)\Big\}.
\label{acaocomplexbse}   
\end{align}
It is easy to see that $m_i$ is related to the mass term of the field $\phi_i$ of the doublet, while $\la_i$ is the coupling constant of the purely quartic self-interaction and $\la_{12}$ denotes the independent coupling constant of the mixed interaction between the complex fields of the doublet. For the theory to recover the original gauge symmetry presented in \eqref{acaocomplex}, we must take the limit in which the scalar parameters degenerate:
\beq
m_1 \,\longrightarrow\,m_2 , \quad \la_1 \,\longrightarrow\,\la , \quad \la_2 \,\longrightarrow\,\la \quad \text{and} \quad \la_{12} \,\longrightarrow\,2\la .
\label{back}
\eeq
\section{One-loop divergences}
\label{sec4}

In this section, we evaluate the one-loop divergences of a deformed
Yang--Mills--Faddeev--Popov (YMFP) theory. We use the background
field method \cite{Abbott1981, Barvinsky2018}, decomposing the
fields as
\begin{align}
    \phi &\to \phi' = \phi+\ph\,,
    \qquad
    \phi^\dagger \to \phi'^\dagger
    = \phi^\dagger+\ph^\dagger\,,
    \qquad
    A_\mu^a \to {A_\mu^a}'=A_\mu^a+B_\mu^a\,,
\end{align}
where $\phi$, $\phi^\dagger$ and $A_\mu^a$ are the background
fields, while $\ph$, $\ph^\dagger$ and $B_\mu^a$ are the
quantum fluctuations.

Our starting point is the gauge-fixed Yang--Mills action,
\begin{align}
    S_{\text{YMFP}}
    =-\frac14\int d^4x\,
    F_{\mu\nu}^aF^{a\mu\nu}
    +S_{\text{gf}}+S_{\text{gh}}\,,
\end{align}
with
\begin{align}
    S_{\text{gf}}
    =-\frac12\int d^4x\,
    \big(\mathcal{D}^{ab}_\mu B^{b\mu}\big)^2\,.
\end{align}
For the one-loop calculation, we need only the part of the
ghost action that is quadratic in the quantum fields,
\begin{align}
    S_{\text{gh}}^{(2)}
    =\int d^4x\,
    \bar c^a\big(\mathcal{D}^2\big)^{ab}c^b\,,
\end{align}
where $c^a$ and $\bar c^a$ are the Faddeev--Popov ghost and
antighost fields, respectively, and $\mathcal{D}_\mu$ is the
background covariant derivative in the adjoint representation. Note that the choice of other gauge-fixing terms and other ghosts would define different deformed models, the equivalence of which is not being assumed.
We define the deformed model by adding the scalar action
for a doublet in the fundamental representation of $SU(2)$,
\begin{align}\label{acaoinicial}
    S=S_{\text{bs}}+S_{\text{YMFP}}\,.
\end{align}
The field decompositions introduced above
are understood in both $S_{\text{bs}}$ and the Yang--Mills
term, with explicit field arguments suppressed. We use
\begin{align}
    F_{\mu\nu}^a
    =\pa_\mu A_\nu^a-\pa_\nu A_\mu^a
    +g\,\ep^{abc}A_\mu^b A_\nu^c\,.
\end{align}

We work in flat spacetime with metric $\eta_{\mu\nu}$.
Repeated indices are summed over, and the appropriate
metric tensors are understood in Lorentz contractions
even when both indices are written in the same position.

The scalar potential explicitly breaks the local $SU(2)$
symmetry of the underlying gauge theory. The gauge-fixing
and ghost terms are retained as part of the definition
of this deformed model. We study the one-loop divergences and the evolution of the couplings within this fixed prescription. Comparison with the results available in the literature for the symmetric
limit provides a consistency check.

We begin by evaluating the corrections to the scalar mass
terms and quartic interactions. To extract the corresponding counterterms, we set the background
gauge field to $A_\mu^a=0$, while retaining the quantum gauge
fluctuations $B_\mu^a$. The diagrams contributing to these terms
are shown in Figures~\ref{fig:mass} and~\ref{fig:lambda}, where
dashed and wavy lines represent the propagators of the quantum
scalar and gauge fields, respectively. The arrows indicate the
charge flow of the complex scalar fields.

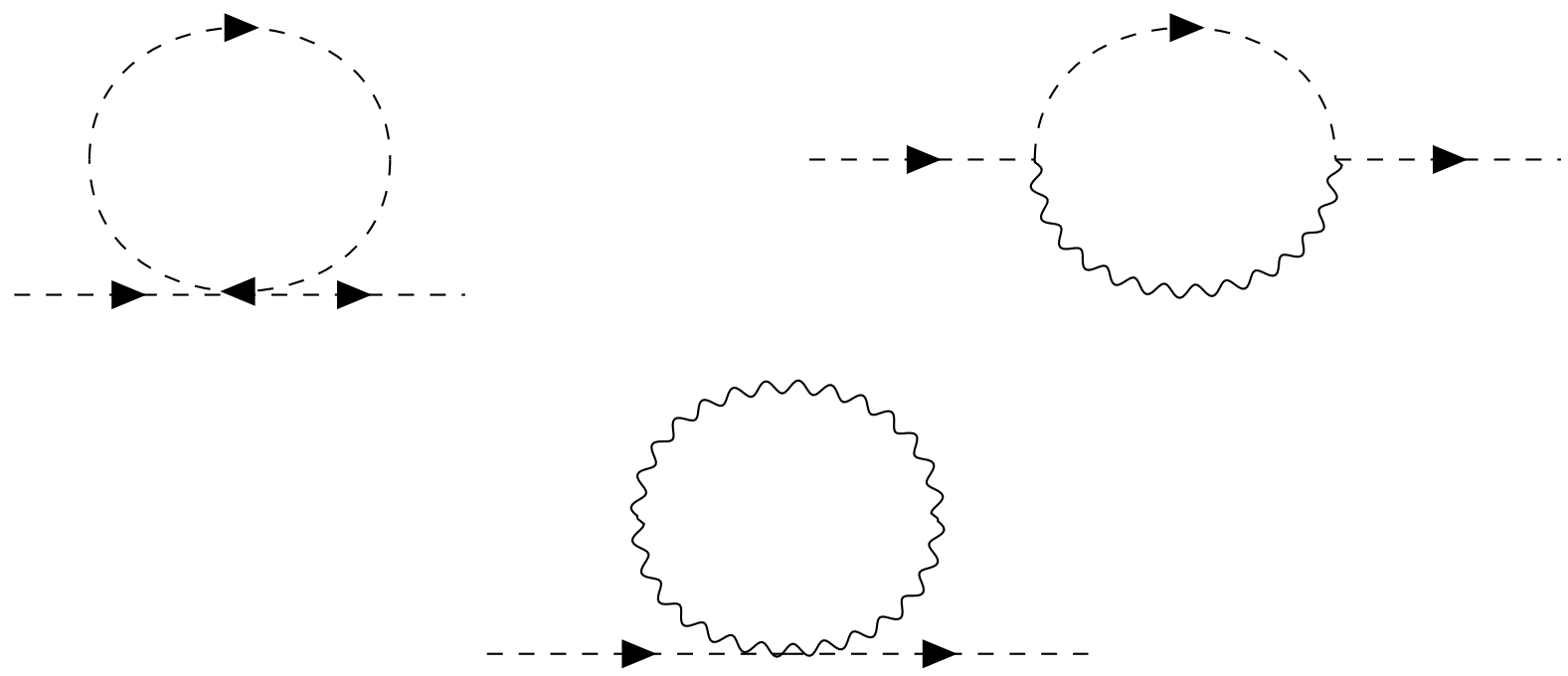
\begin{figure}[H]
\begin{center}
\begin{tikzpicture}
\begin{feynman}
\vertex (v1) at (0,-0.9);
\vertex (v2) at (-1,0);
\vertex (v3) at (1,0);
\vertex (a1) at (-1.5,-0.9);
\vertex (a2) at (1.5,-0.9);

\diagram* {
(a1) -- [charged scalar] (v1),
(v1) -- [charged scalar] (a2),
(v2) -- [charged scalar, half left] (v3),
(v3) -- [charged scalar, half left] (v2),
};
\end{feynman}
\end{tikzpicture}
\hspace{2cm}
\begin{tikzpicture}
\begin{feynman}

\vertex (a1) at (-1.5,-0.5);
\vertex [dot] (v1) at (0,-0.5);

\vertex [dot] (v2) at (2,-0.5);
\vertex (out1) at (3.5,-0.5);

\diagram* {
(a1) -- [charged scalar] (v1),

(v1) -- [charged scalar, half left] (v2),
(v1) -- [photon, half right] (v2),

(v2) -- [charged scalar] (out1),
};

\end{feynman}
\end{tikzpicture}
\hspace{2cm}
\begin{tikzpicture}
\begin{feynman}
\vertex (v1) at (0,-0.9);
\vertex (v2) at (-1,0);
\vertex (v3) at (1,0);
\vertex (a1) at (-2,-0.9);
\vertex (a2) at (2,-0.9);

\diagram* {
(a1) -- [charged scalar] (v1),
(v1) -- [charged scalar] (a2),
(v2) -- [photon, half left] (v3),
(v3) -- [photon, half left] (v2),
};
\end{feynman}
\end{tikzpicture}
\end{center}
\caption{Diagrams contributing to the mass term.}
\label{fig:mass}
\end{figure}
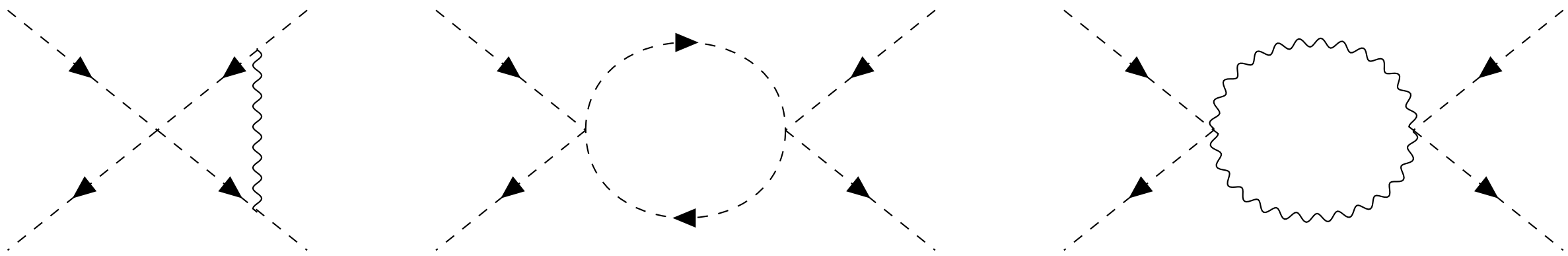
\begin{figure}[H]
\begin{center}
\begin{tikzpicture}
  \begin{feynman}
    \vertex (a1) at (-1.5,1.2);
    \vertex (a2) at (-1.5,-1.2);
    \vertex (a3) at (1.5,1.2);
    \vertex (a4) at (1.5,-1.2);
    \vertex (a5) at (1.0,0.82);
    \vertex (a6) at (1.0,-0.82);
    \vertex (v1) at (0,0);
    
    \diagram* {
      (a1) --[charged scalar] (v1),
      (v1) --[charged scalar] (a2),
      (a3) --[charged scalar] (v1),
      (v1) -- [charged scalar](a4),
      (a5) -- [photon] (a6),
    };
  \end{feynman}
\end{tikzpicture}
\hspace{1cm}
\begin{tikzpicture}
    \begin{feynman}
      \vertex (a1) at (-1.5,1.2);
    \vertex (a2) at (-1.5,-1.2);
    \vertex (a3) at (3.5,1.2);
    \vertex (a4) at (3.5,-1.2);
    \vertex (v1) at (0,0);
    \vertex (v2) at (2,0);

    \diagram* {
      (a1) --[charged scalar] (v1),
      (v1) --[charged scalar] (a2),
      (a3) --[charged scalar] (v2),
      (v2) -- [charged scalar](a4),
      (v1) -- [charged scalar,half left] (v2),
      (v2) -- [charged scalar,half left] (v1)
    };
    \end{feynman}
\end{tikzpicture}
\hspace{1cm}
\begin{tikzpicture}
    \begin{feynman}
      \vertex (a1) at (-1.5,1.2);
    \vertex (a2) at (-1.5,-1.2);
    \vertex (a3) at (3.5,1.2);
    \vertex (a4) at (3.5,-1.2);
    \vertex (v1) at (0,0);
    \vertex (v2) at (2,0);

    \diagram* {
      (a1) --[charged scalar] (v1),
      (v1) --[charged scalar] (a2),
      (a3) --[charged scalar] (v2),
      (v2) -- [charged scalar](a4),
      (v1) -- [photon,half left] (v2),
      (v2) -- [photon,half left] (v1)
    };
    \end{feynman}
\end{tikzpicture}
\end{center}
\caption{Diagrams contributing to the quartic term.}
\label{fig:lambda}
\end{figure}
Note that by expanding $\phi^\dagger \phi$ in terms of the complex scalar fields $\phi_1$ and $\phi_2$, each diagram can be decomposed into other diagrams of the same type, considering propagators of the type $\phi_1^\ast \phi_1$ and $\phi_2^\ast \phi_2$.
Setting the background gauge field to zero and retaining only the terms quadratic in the quantum gauge fluctuations $B_\mu^a$, the Yang--Mills action becomes
\begin{align}
S_{\text{YM}}^{(2)}[A=0]
&= -\frac{1}{4}\int d^4x\,
\left(\partial_\mu B_\nu^a-\partial_\nu B_\mu^a\right)^2
\nonumber\\
&= \frac{1}{2}\int d^4x\,
B^{\mu a}
\left(\delta_\mu^\nu\Box-\partial_\mu\partial^\nu\right)
B_\nu^a\,,
\label{acao}
\end{align}
where
\begin{align}
    \Box=\eta^{\mu\nu}\partial_\mu\partial_\nu.
\end{align}
For a vanishing background field $A_\mu^a$, the quadratic part of the gauge-fixing action in the quantum fluctuations is given by
\begin{align}
   S_{gf} = - \,\frac{1}{2}\int d^4x \big(\pa_\mu B^{\mu a}\big)^2.
\end{align}
 To simplify the calculation, we temporarily impose the restriction $\lambda_{12}=2\sqrt{\lambda_1\lambda_2}$. This allows us to express the quartic scalar potential in the compact form $(\phi^\dagger\Lambda\phi)^2$, with $\Lambda=\operatorname{diag}(\sqrt{\lambda_1},\sqrt{\lambda_2})$, considerably simplifying the Hessian matrix and the evaluation of the traces in the Schwinger--DeWitt technique. The extension to an independent mixed coupling is justified in Appendix B. Defining the mass matrix $M=\diag(m_1^2,m_2^2)$, the scalar
action can be written as
\begin{align}
    S_{\text{bs}} = \int d^4x \Big\{
    (D^\mu\phi)^\dagger(D_\mu\phi)
    - \phi^\dagger M\phi
    - (\phi^\dagger\La\phi)^2
    \Big\}.
\end{align}
Therefore, using the background field method, the bilinear form in the quantum fields is given by
\begin{align}
    H[\phi, \phi^\dagger, A_\mu^a = 0] = \begin{pmatrix}
        H_{\phi\phi^\dagger} &H_{\phi\phi} & H_{\phi B^b_\nu}\\
        H_{\phi^\dagger\phi^\dagger} &  H_{\phi^\dagger\phi} & H_{\phi^\dagger B^b_\nu} \\
         H_{B^a_\mu \phi^\dagger} &  H_{B^a_\mu \phi}&H_{B^a_\mu B^b_\nu}
    \end{pmatrix}.\label{hessiana}
\end{align}
Here, the entries of \eqref{hessiana} are:
\begin{align}
    H_{\phi\phi^\dagger} &= \delta^{ab} \Box + M_{ab} + 2(\phi^\dagger \La \phi)\La_{ba} +2(\phi^{\dagger}\La)_a (\La\phi)_b\,,\nn\\
    H_{\phi\phi} &= 2(\phi^{\dagger}\La)_a (\phi^{\dagger}\La)_b\,, \nn\\
    H_{\phi B^b_\nu} &= \frac{i}{2}g(\phi^\dagger t^b)_a  \pa_\nu + ig(\pa_\nu \phi^\dagger t^b)_a\,,\nn\\
    H_{\phi^\dagger\phi^\dagger} &=2(\La\phi)_a (\La\phi)_b \,, \nn\\
    H_{\phi^\dagger\phi} &= \delta^{ab} \Box + M_{ab} + 2(\phi^\dagger \La \phi)\La_{ab} +2(\phi^{\dagger}\La)_b (\La\phi)_a\,,\nn\\
    H_{\phi^\dagger B^b_\nu} &= -\frac{i}{2}g(t^b \phi)_a  \pa_\nu - ig(t^b\pa_\nu \phi)_a\,, \nn\\
    H_{B^a_\mu \phi^\dagger} &= -\frac{i}{2}g(t^a \phi)_b  \pa_\mu + \frac{i}{2}g(t^a\pa_\mu \phi)_b \,,\nn\\
    H_{B^a_\mu \phi} &= \frac{i}{2}g(\phi^\dagger t^a)_b  \pa_\mu -\frac{i}{2}g(\pa_\mu \phi^\dagger t^a)_b\,,\nn\\
    H_{B^a_\mu B^b_\nu}
&= \delta^{ab}\eta_{\mu\nu}\Box
    + \frac{g^2}{2}(\phi^\dagger\phi)
      \delta^{ab}\eta_{\mu\nu}
\,.
\end{align}
At this point, it is worth noting that the Kronecker deltas $\delta_{ab}$ in the scalar and vector sectors play different roles. In the scalar sector, the indices run from $a,b = 1,2$, identifying each complex field of the doublet. For the vector sector, however, the indices run from $a,b = 1,2,3$, as they are related to the indices of the $SU(2)$ algebra generators.

It is straightforward to see that we can rewrite the bilinear form in the quantum fields as
\begin{align}
    \hat{H} = \hat{1} \Box + 2 \hat{h}^\al \pa_\al + \hat{\Pi}, 
\end{align}
where $\hat{1} = \diag(\delta_{ab}, \delta_{ab} , \delta_{ab}\eta_{\mu\nu})$,  
\begin{align}
    \hat{h}^\al = (\hat{h}^{\al}_{ab})_{\mu\nu} =  \begin{pmatrix}
        0&0& \frac{i}{4}g(\phi^\dagger t^b)_a \delta^{\al}_\nu\\
        0 &0 & -\frac{i}{4} g(t^b \phi)_a \delta^{\al}_\nu\\
        -\frac{i}{4} g(t^a \phi)_b \delta^{\al}_\mu &  \frac{i}{4}g(\phi^\dagger t^a)_b \delta^{\al}_\mu & 0
    \end{pmatrix}
\end{align}
and the elements of the operator $\hat{\Pi} = (\hat{\Pi}_{ab})_{\mu\nu}$ are given by
\begin{align}
    \hat{\Pi}_{11} &= M_{ab} + 2(\phi^\dagger \La \phi)\La_{ba} +2(\phi^{\dagger}\La)_a (\La\phi)_b \,, \nn\\
    \hat{\Pi}_{12} &= 2(\phi^{\dagger}\La)_a (\phi^{\dagger}\La)_b \,, \nn\\
    \hat{\Pi}_{13} &= ig(\pa_\nu \phi^\dagger t^b)_a \,, \nn\\
    \hat{\Pi}_{21} &= 2(\La\phi)_a (\La\phi)_b \,, \nn\\
    \hat{\Pi}_{22} &= M_{ab} + 2(\phi^\dagger \La \phi)\La_{ab} +2(\phi^{\dagger}\La)_b (\La\phi)_a \,, \nn\\
    \hat{\Pi}_{23} &= - ig(t^b\pa_\nu \phi)_a \,, \nn\\
    \hat{\Pi}_{31} &= \frac{i}{2}g(t^a\pa_\mu \phi)_b \,, \nn\\
    \hat{\Pi}_{32} &= -\frac{i}{2}g(\pa_\mu \phi^\dagger t^a)_b \,, \nn\\
    \hat{\Pi}_{33} &= 
     \frac{g^2}{2}(\phi^\dagger\phi)
      \delta^{ab}\eta_{\mu\nu}
 \,.
\end{align}
The one-loop contribution to the effective action is given by
\beq
    \bar{\Ga}^{(1)}_{div}
    \, =\,
    \frac{i}{2} \Tr \log \hat{H}'
    \, -\,
    i \Tr \log \hat{H}_{gh} ,
\eeq
where $\hat{H}_{gh}$ is the operator corresponding to the ghost action,
\beq
\hat{H}_{gh}\,=\, \Box \, .
\eeq
An important point is that, by setting the background gauge field $A_\mu^a$ to zero, the ghost term yields no one-loop contributions to the mass and quartic terms; therefore, we do not need to consider it for this purpose. This occurs because Yang-Mills ghosts couple exclusively to gauge fields; in the absence of external gauge fields, it is impossible to construct a connected one-loop diagram with external scalar legs and an internal ghost loop.

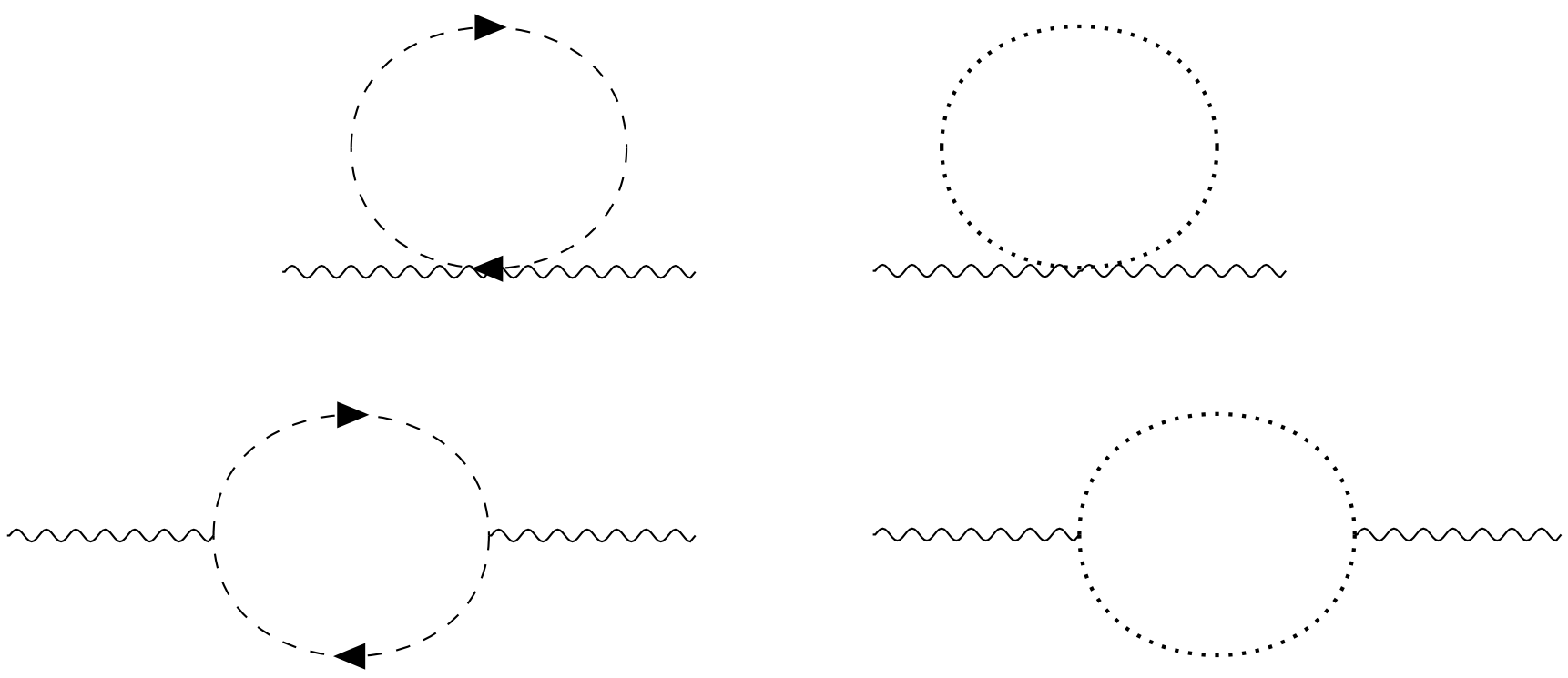
\begin{figure}[H]
   \begin{center}
\begin{tikzpicture}
\begin{feynman}
\vertex (v1) at (0,-0.9);
\vertex (v2) at (-1,0);
\vertex (v3) at (1,0);
\vertex (a1) at (-1.5,-0.9);
\vertex (a2) at (1.5,-0.9);

\diagram* {
(a1) -- [photon] (v1),
(v1) -- [photon] (a2),
(v2) -- [charged scalar, half left] (v3),
(v3) -- [charged scalar, half left] (v2),
};
\end{feynman}
\end{tikzpicture}
\hspace{1cm}
\begin{tikzpicture}
\begin{feynman}
\vertex (v1) at (0,-0.9);
\vertex (v2) at (-1,0);
\vertex (v3) at (1,0);
\vertex (a1) at (-1.5,-0.9);
\vertex (a2) at (1.5,-0.9);

\diagram* {
(a1) -- [photon] (v1),
(v1) -- [photon] (a2),
(v2) -- [ghost, half left] (v3),
(v3) -- [ghost, half left] (v2),
};
\end{feynman}
\end{tikzpicture}
\end{center}
\vspace{-0.5cm}
\begin{center}
\begin{tikzpicture}
\begin{feynman}

\vertex (a1) at (-1.5,-0.5);
\vertex [dot] (v1) at (0,-0.5);

\vertex [dot] (v2) at (2,-0.5);
\vertex (out1) at (3.5,-0.5);

\diagram* {
(a1) -- [photon] (v1),

(v1) -- [charged scalar, half left] (v2),
(v2) -- [charged scalar, half left] (v1),

(v2) -- [photon] (out1),
};

\end{feynman}
\end{tikzpicture}
\hspace{1cm}
\begin{tikzpicture}
\begin{feynman}
\vertex (a1) at (-1.5,-0.5);
\vertex [dot] (v1) at (0,-0.5);

\vertex [dot] (v2) at (2,-0.5);
\vertex (out1) at (3.5,-0.5);

\diagram* {
(a1) -- [photon] (v1),

(v1) -- [ghost, half left] (v2),
(v1) -- [ghost, half right] (v2),

(v2) -- [photon] (out1),
};
\end{feynman}
\end{tikzpicture}
\end{center}
    \caption{Diagrams contributing to $F_{\mu\nu}^aF_{\mu\nu}^a$.}
    \label{fig:F^2}
\end{figure}

Using the Schwinger-DeWitt technique, the divergent part of the one-loop effective action is given by \cite{DeWitt1965, DeWitt2003}
\beq
&&
\bar{\Ga}^{(1)}_{div}
\,\,=\,\,
-\, \frac{\mu^{n-4}}{\vp}\,\int d^n x\,
\,\tr \bigg\{\,
\frac{1}{2}\,\hat{P}^2 + \frac{1}{12}\,\hat{\mathcal{S}}_{\al\be}^2
+ \frac{1}{6}\,\Box \hat{P}
\bigg\},
\label{tracoa2}
\eeq
where $\mu$ is the renormalization scale and $\vp = (4\pi)^2(n-4)$ is the dimensional regularization parameter. The operators $\hat{P}$ and $\hat{\mathcal{S}}_{\al\be}$ in (\ref{tracoa2}) are given by
\beq
&&
\hat{P}
\,=\,
\hat{\Pi} - \pa_\al \hat{h}^\al - \hat{h}_\al\hat{h}^\al,
\label{eqP}
\eeq
\beq
&&
\hat{\mathcal{S}}_{\al\be}
\,=\,
\hat{\mathcal{R}}_{\be\al}
+ \pa_\be\hat{h}_\al - \pa_\al\hat{h}_\be
+ \hat{h}_\be \hat{h}_\al - \hat{h}_\al \hat{h}_\be,
\label{eqS}
\eeq
where $\hat{\mathcal{R}}_{\al\be} = 0$ in both the scalar and vector sectors, since we are working with commuting partial derivatives rather than covariant ones, as would be the case in curved spacetime.
Therefore, the divergent part of the one-loop effective action is given by
\begin{align}
    \Ga^{(1)}_{\text{div}}[A_\mu^a = 0] =& -\frac{\mu^{n-4}}{\ep} \int d^n x \bigg\{ -3g^2 (\pa_\la \phi^\dagger)(\pa_\la \phi) + (\tr M^2) \nn\\
    &-g^2 (\tr M) (\phi^\dagger \phi) + \frac{g^2}{2}\phi^\dagger M \phi + 4\, (\tr M\La)(\phi^\dagger \La \phi) + 4(\phi^\dagger M \La^2 \phi) \nn\\
    & +\, {\frac{9}{8}}g^4(\phi^\dagger \phi)^2 + \left[4(\tr \La^2) + 3g^2\right] (\phi^\dagger \La \phi)^2 + 8(\phi^\dagger \La^2 \phi)^2 \nn\\
    &- 2g^2 (\tr \La)(\phi^\dagger \La \phi)(\phi^\dagger \phi) + 8(\phi^\dagger \La \phi)(\phi^\dagger \La^3 \phi)-2g^2 (\phi^\dagger \phi)(\phi^\dagger \La^2 \phi)\bigg\}.
\end{align}
Notice that, at first glance, some quartic terms that are not present in the original classical action seem to appear. However, by rewriting the matrix expression above in terms of each complex field of the doublet, we obtain
\begin{align}\label{oneloop1}
   \Ga^{(1)}_{\text{div}}[A_\mu^a = 0] =& -\frac{\mu^{n-4}}{\ep} \int d^n x \bigg\{ -3g^2(\pa_\mu\phi_1^*)(\pa_\mu\phi_1) -3g^2(\pa_\mu\phi_2^*)(\pa_\mu\phi_2) + (\tr M^2) \nn\\
   &+\,\left[-\frac12 m_1^2g^2 - m_2^2g^2 + 8m_1^2\la_1 + 4m_2^2\sqrt{\la_1\la_2}\right](\phi_1^*\phi_1)\nn\\
   &+\,\left[-\frac12 m_2^2g^2 - m_1^2g^2 + 8m_2^2\la_2 + 4m_1^2\sqrt{\la_1\la_2}\right](\phi_2^*\phi_2)\nn\\
   & +\,\left[{\frac{9}{8}}g^4 - \la_1 g^2 - 2\sqrt{\la_1\la_2}g^2 + 20\la_1^2 + 4\la_1\la_2\right](\phi_1^*\phi_1)^2\nn\\
   &+\,\left[{\frac{9}{8}}g^4 - \la_2 g^2 - 2\sqrt{\la_1\la_2}g^2 + 20\la_2^2 + 4\la_1\la_2\right](\phi_2^*\phi_2)^2\nn\\
   &\hspace{-2.70cm}+\,\left[{\frac{9}{4}}g^4+2g^2\sqrt{\la_1\la_2} - 4g^2(\la_1+\la_2) + 16(\la_1 + \la_2)\sqrt{\la_1\la_2}+16\la_1\la_2\right](\phi_1^*\phi_1)(\phi_2^*\phi_2)\bigg\},
\end{align}
that is, by comparing with \eqref{acaocomplexbse}, we can see that the scalar terms are renormalizable. {The reconstruction of the covariant scalar kinetic counterterm for a non-zero background gauge field is examined in Appendix A.}
Now, to obtain the corrections to the quadratic term $F_{\mu\nu}^aF_{\mu\nu}^a$, we can consider the action
\begin{align}\label{acaosimpl}
    S_{A} = \int d^4x \left\{(D_\mu \phi)^\dagger(D_\mu \phi) - \frac14 F_{\mu\nu}^aF_{\mu\nu}^a \right\}.
\end{align}
Furthermore, when employing the background field method, we can set the background of the complex scalar fields to zero, since the contributions to $F_{\mu\nu}^aF_{\mu\nu}^a$ arise from the diagrams in Figure \eqref{fig:F^2}.
Here, the dashed lines still represent $\phi^\dagger\phi$ propagators, while the wavy lines are associated with $A_\mu^a A_\nu^b$ propagators. The dotted lines correspond to the Faddeev--Popov ghost field propagators, $\bar c c$.
Thus, using the background field method, we can show that 
\begin{align}\label{Fmunu}
    A_\mu \to A_\mu' = A_\mu + B_\mu \implies F_{\mu\nu}^a[A_\mu'] = F_{\mu\nu}^a[A_\mu] + (\mathcal{D}_\mu B_\nu)^a - (\mathcal{D}_\nu B_\mu)^a + g\ep^{abc}B_\mu^bB_\nu^c ,
\end{align}
where 
\begin{align}\label{adj}
    (\mathcal{D}_\mu B_\nu)^a = \mathcal{D}_\mu^{ab} B_\nu^b = \left(\delta^{ab} \pa_\mu + g \ep^{acb}A_\mu^c\right)B^{b}_\nu\,.
\end{align}
Using \eqref{Fmunu} and \eqref{adj}, we can obtain the quadratic form in the quantum fields for the vector term of the action \eqref{acaosimpl},
\begin{align}
    -\frac{1}{4} F_{\mu\nu}^{a}F_{\mu\nu}^a &= -\frac12 \left[2g\bar{F}_{\mu\nu}^{a}\ep^{abc}B_\mu^b B_\nu^c - B_\nu^a (\mathcal{D}^2)^{ab} B_\nu^b + B^a_\mu (\mathcal{D}_\mu\mathcal{D}_\nu)^{ab} B_\nu^b\right]\nn\\
    &=B_{\mu}^a \left[\frac12 \delta_{\mu\nu}(\mathcal{D}^2)^{ab} - \frac12(\mathcal{D}_\mu\mathcal{D}_\nu)^{ab} -g\bar{F}_{\mu\nu}^{c}\ep^{abc}  \right]B_{\nu}^b\,,
\end{align}
where $\bar{F}_{\mu\nu}^{c}$ is the tensor ${F}_{\mu\nu}^{c}$ constructed from the background gauge field $A_\mu^a$. As for the scalar kinetic term, since we set the scalar background field to zero, 
\begin{align}
    \phi \to \phi' = 0 + \ph \implies(D_\mu\phi')^\dagger(D_\mu\phi') &= (D_\mu\ph)^\dagger(D_\mu\ph)\nn\\
    &= \ph^\dagger_a\left[-\delta^{ab} D^2\right]\ph_b\,.
\end{align}
Taking into account the gauge-fixing term, 
\begin{align}
    S_{gf} = -\frac12 \int d^4x \left(\mathcal{D}_\mu^{ab} B_\mu^b\right)^2 = \frac{1}{2} \int d^4x \,B_\mu^a \left(\mathcal{D_\mu}\mathcal{D_\nu}\right)^{ab}B_\nu^b,
\end{align}
the Hessian can be written as:
\begin{align}
    S^{(2)} = \int d^4x  \begin{pmatrix}
        \ph^\dagger_a & B_\mu^a 
    \end{pmatrix} \hat{H'}\begin{pmatrix}
        \ph_b\\
        B_{\nu}^b
    \end{pmatrix}\,.
\end{align}
Here, 
\begin{align}
    \hat{H'} = \begin{pmatrix}
        -1 & 0 \\
        0 & \frac12
    \end{pmatrix}\begin{pmatrix}
        \delta^{ab} D^2 & 0 \\
        0 & \delta_{\mu\nu} \mathcal{D}^2 - 2g\bar{F}^c_{\mu\nu} \ep^{abc}
    \end{pmatrix}\,.
\end{align}
The one-loop contribution is given by
\begin{align}\label{1loopA}
    \bar \Ga^{(1)}_{\text{div}}[\phi = 0] \,=\, \frac{i}{2} \Tr \log \begin{pmatrix}
        \delta^{ab} D^2 & 0 \\
        0 & \delta_{\mu\nu} \mathcal{D}^2 - 2g\bar{F}^c_{\mu\nu} \ep^{abc}
    \end{pmatrix} - i\Tr \log \hat{H}_{\text{gh}},
\end{align}
where $\hat{H}_{\text{gh}}$ is the ghost operator,
\begin{align}
    \hat{H}_{\text{gh}} = (\mathcal{D}^2)^{ab}\,.
\end{align}
For the ghost operator, the only term that contributes to the divergences is the commutator of the covariant derivatives. It is well known that
\begin{align}
    [\mathcal{D_\mu}, \mathcal{D_\nu}] = g \ep^{acb} {F}_{\mu\nu}^c,
\end{align}
and thus,
\begin{align}\label{ghcontri}
    - i\Tr \log \hat{H}_{\text{gh}} &= -\frac{\mu^{n-4}}{\ep} \int d^n x \str\left\{ \frac{1}{12}\left[g \ep^{acb} {F}_{\mu\nu}^c\right]^2\right\}\nn\\
    &= -\frac{\mu^{n-4}}{\ep} \int d^n x \left\{ \frac{1}{3}g^2  ({F}_{\mu\nu}^a)^2\right\}.
\end{align}
For the remaining term in Eq.\,\eqref{1loopA}, we can once again use the Schwinger-DeWitt technique. In this case, 
\begin{align}
    \hat{1} = \diag(\delta^{ab},\delta_{\mu\nu}) \,\,\text{ and } \,\, \hat{P} = \begin{pmatrix}
        0 & 0 \\
        0 &  - 2g\bar{F}^c_{\mu\nu} \ep^{abc}
    \end{pmatrix}\,.\label{eqP2}
\end{align}
The operator $\hat{S}_{\al\be}$ is given by 
\begin{align}
    \hat{S}_{\al\be} = \begin{pmatrix}
        -\frac{i}{2}gF_{\al\be}^a t^a & 0 \\
        0 & g \ep^{acb} F_{\al\be}^c
    \end{pmatrix}\,.\label{eqS2}
\end{align}
From Equations \eqref{eqP2} and \eqref{eqS2}, we can obtain the one-loop contribution:
\begin{align}
    \frac{i}{2} \Tr \log \begin{pmatrix}
        \delta^{ab} D^2 & 0 \\
        0 & \delta_{\mu\nu} \mathcal{D}^2 - 2g\bar{F}^c_{\mu\nu} \ep^{abc}
    \end{pmatrix} &= -\frac{\mu^{n-4}}{\ep} \int d^n x \tr  \left\{\frac12 \hat{P}^2 + \frac{1}{12}\hat{S}_{\al\be}^2\right\}\nn\\
    & = -\frac{\mu^{n-4}}{\ep} \int d^n x   \left\{4g^2{(F_{\mu\nu}^a)}^2 - \frac34g^2{(F_{\mu\nu}^a)}^2\right\}\nn\\
    & = -\frac{\mu^{n-4}}{\ep} \int d^n x   \left\{\frac{13}{4}g^2{(F_{\mu\nu}^a)}^2\right\}\,.
\end{align}
Therefore, including the ghost contribution, we obtain
\begin{align}\label{oneloop2}
    \bar \Ga^{(1)}_{\text{div}}[\phi = 0] \,\,=\,\, -\frac{\mu^{n-4}}{\ep} \int d^n x   \left\{\frac{43}{12}g^2{(F_{\mu\nu}^a)}^2 \right\}.
\end{align}
We can observe that by computing only the contributions from the vector sector, we recover the well-known result \cite{Buchbinder2021},
\begin{align}
     \bar \Ga^{(1)}_{\text{div}}[\phi = 0]_{\text{vector}} \,\,=\,\, -\frac{\mu^{n-4}}{\ep} \int d^n x   \left\{\frac{11}{3}g^2{(F_{\mu\nu}^a)}^2 \right\}\,.
\end{align}
It is important to observe that the explicit symmetry breaking introduced in the scalar sector does not propagate {to the UV divergences of the pure Yang--Mills sector} at the one-loop level. This isolation occurs for {three} fundamental reasons. First, from a diagrammatic point of view, the vacuum polarization graphs that generate the one-loop contributions to the pure gauge sector are completely independent of the symmetry-breaking quartic scalar vertices ($\lambda_i$). 
Second, due to dimensional constraints dictated by power-counting renormalizability, we cannot radiatively generate {UV-divergent} mass-dependent corrections to the gauge kinetic term, such as terms proportional to $(m_1^2-m_2^2)(F_{\mu\nu}^a)^2$. Such terms would correspond to non-renormalizable dimension-6 operators. {Third, as in Appendix A, the relevant heat-kernel coefficient contains $\hat U_s^2$ and $\hat S_{\mu\nu}^2$, with no mixed product. For a vanishing scalar background, $\hat U_s=M$ contributes only to mass-dependent vacuum terms, while $\hat S_{\mu\nu}^2$ generates the $F_{\mu\nu}^aF^{a\mu\nu}$ counterterm. Thus, no mixed UV-divergent terms schematically of the form $M^i A^j$, with $i,j>0$, are generated at one loop. In particular, this excludes gauge-field mass counterterms, which would otherwise be allowed by power counting.} 

Consequently, the one-loop divergences of the vector sector remain completely independent of the scalar symmetry-breaking parameters. From Equations \eqref{oneloop1} and \eqref{oneloop2}, {together with the dimensional and diagrammatic analysis presented in Appendix A, we conclude that} counterterms of the form
\begin{align}
    \pa_\mu \phi^\dagger \pa_\mu\phi\,,
\end{align}
actually correspond, when we do not employ the simplification of setting the background fields to zero, to
\begin{align}
    (D_\mu \phi)^\dagger (D_\mu \phi)\,.
\end{align}
With this argument, the final form of the one-loop corrections for this theory is given by
\begin{align}
     \bar{\Ga}^{(1)}_{\text{div}} =& -\frac{\mu^{n-4}}{\ep} \int d^n x \bigg\{ -3g^2(D_\mu\phi_1^*)(D_\mu\phi_1) -3g^2(D_\mu\phi_2^*)(D_\mu\phi_2)  \nn\\
   &+\,\left[-\frac12 m_1^2g^2 - m_2^2g^2 + 8m_1^2\la_1 + 4m_2^2\sqrt{\la_1\la_2}\right](\phi_1^*\phi_1)\nn\\
   &+\,\left[-\frac12 m_2^2g^2 - m_1^2g^2 + 8m_2^2\la_2 + 4m_1^2\sqrt{\la_1\la_2}\right](\phi_2^*\phi_2)\nn\\
   & +\,\left[{\frac{9}{8}}g^4 - \la_1 g^2 - 2\sqrt{\la_1\la_2}g^2 + 20\la_1^2 + 4\la_1\la_2\right](\phi_1^*\phi_1)^2\nn\\
   &\hspace{-1cm}+\,\left[{\frac{9}{8}}g^4 - \la_2 g^2 - 2\sqrt{\la_1\la_2}g^2 + 20\la_2^2 + 4\la_1\la_2\right](\phi_2^*\phi_2)^2 + \frac{43}{12}g^2{(F_{\mu\nu}^a)}^2\nn\\
   &\hspace{-2.3cm}+\,\left[{\frac{9}{4}}g^4+2g^2\sqrt{\la_1\la_2} - 4g^2(\la_1+\la_2) + 16(\la_1 + \la_2)\sqrt{\la_1\la_2}+16\la_1\la_2\right](\phi_1^*\phi_1)(\phi_2^*\phi_2) \bigg\}\,\nn\\
   &\hspace{9cm} + \bar{\Ga}^{(1)}_{\text{div},\,\text{vac}}\,.
\end{align}

The one-loop counterterms preserve the structure of the classical action in the sectors considered, consistently with the renormalizability of the deformed model at one-loop order.

Although the calculation was performed under the temporary restriction $\lambda_{12}=2\sqrt{\lambda_1\lambda_2}$, the extension to an independent mixed coupling can be justified using the polynomial structure of the one-loop divergences and the corresponding Feynman diagrams, as shown in Appendix B.

The Feynman rules of the unrestricted theory generate only integer powers of the independent couplings. Therefore, terms proportional to $2\sqrt{\lambda_1\lambda_2}$ in the restricted calculation can be identified with contributions proportional to $\lambda_{12}$. For terms proportional to $4\lambda_1\lambda_2$, the restriction alone does not distinguish contributions proportional to $\lambda_{12}^2$ from those proportional to $\lambda_1\lambda_2$. However,
the diagrammatic analysis in Appendix B excludes the latter contributions, justifying the substitution $4\lambda_1\lambda_2\longrightarrow\lambda_{12}^2$.

With this argument, the final form of the one-loop corrections for our generalized theory is given by:
\begin{align}
\bar{\Gamma}^{(1)}_{div} &= -\frac{\mu^{n-4}}{\epsilon} \int d^nx \Bigg\{ -3g^2(D_\mu \phi_1^*)(D_\mu \phi_1) - 3g^2(D_\mu \phi_2^*)(D_\mu \phi_2) \nonumber \\
&\quad + \left[ -\frac{1}{2}m_1^2 g^2 - m_2^2 g^2 + 8m_1^2\la_1 + 2m_2^2\la_{12} \right] (\phi_1^*\phi_1) \nonumber \\
&\quad + \left[ -\frac{1}{2}m_2^2 g^2 - m_1^2 g^2 + 8m_2^2\la_2 + 2m_1^2\la_{12} \right] (\phi_2^*\phi_2) \nonumber \\
&\quad + \left[ {\frac{9}{8}}g^4 - \la_1 g^2 - \la_{12} g^2 + 20\la_1^2 + \la_{12}^2 \right] (\phi_1^*\phi_1)^2 \nonumber \\
&\quad + \left[ {\frac{9}{8}}g^4 - \la_2 g^2 - \la_{12} g^2 + 20\la_2^2 + \la_{12}^2 \right] (\phi_2^*\phi_2)^2 \nonumber \\
&\quad + \frac{43}{12}g^2(F^a_{\mu\nu})^2 + \left[ {\frac{9}{4}}g^4 + g^2\la_{12} - 4g^2(\la_1+\la_2) + 8(\la_1+\la_2)\la_{12} + 4\la_{12}^2 \right] (\phi_1^*\phi_1)(\phi_2^*\phi_2) \Bigg\} \nonumber \\
&\quad + \bar{\Gamma}^{(1)}_{div, vac} \,.
\label{eq:acao_efetiva_final}
\end{align}

{ Equation \eqref{eq:acao_efetiva_final} gives the one-loop divergences in the bosonic background sector, with vanishing ghost backgrounds. The resulting counterterms preserve the structure of the bosonic action considered here. Having extracted these divergences, the next natural step is to determine the associated beta functions to analyze the ultraviolet behavior of the theory.}
\section{One-loop Renormalization group equations}

With the one-loop corrections at hand, we are now in a position to study the behavior of the couplings with respect to the energy scale. The unrenormalized (bare) action $S_{0}$ can be decomposed as $S_{0} = S + \Delta S$, where $S$ represents the renormalized tree-level action, and $\Delta S$ corresponds to the counterterms that cancel the divergences, defined at one-loop order by $\Delta S = - \bar{\Ga}^{(1)}_{\text{div}}$. {The bosonic part of the bare action, relevant for the renormalization of the scalar and gauge couplings considered here, is given by}
\begin{align}
    S_{0} \,\,&=\,\, 
\int d^4x \Big\{(D_\mu\phi_1)^* \,(D_\mu\phi_1) 
+(D_\mu\phi_2)^* \,(D_\mu\phi_2)- m_1^2\phi^*_1\phi_1 - m_2^2\phi^*_2\phi_2\nn\\
&\quad- \la_1(\phi^*_1\phi_1)^2- \la_2(\phi^*_2\phi_2)^2- \la_{12}(\phi^*_1\phi_1)(\phi^*_2\phi_2) -\frac14 (F_{\mu\nu}^a)^2\Big\},
\end{align}
where it is implicitly understood that all fields, couplings, and masses appearing inside the integrand are the bare (unrenormalized) quantities, traditionally denoted with a zero index. Explicitly, the bare covariant derivative is defined as $D_\mu^0 = \partial_\mu - i g_0 t^a (A_\mu^a)_0 / 2$, which acts on the bare scalar multiplets. The connection between these bare quantities and their renormalized counterparts is established through the standard multiplicative field renormalization relations:
\begin{align}
    (\phi_i^*)_0 &= \mu^{\frac{n-4}{2}} \left(1-\frac{3g^2}{2\ep}\right)\phi_{i}^*\,, \nn\\
    (\phi_i)_0 &= \mu^{\frac{n-4}{2}} \left(1-\frac{3g^2}{2\ep}\right)\phi_{i}\,,\nn\\
    (A_\mu^a)_0 &= \mu^{\frac{n-4}{2}} \left(1 - \frac{43}{6}\frac{g^2}{\ep}\right)A_\mu^a\,.
\end{align}
{Matching the quadratic, cubic and quartic parts of the calculated
$F_{\mu\nu}^aF^{a\mu\nu}$ counterterm with the bare Yang--Mills
term shows that the product of the gauge coupling and the background
field is unchanged by renormalization at one loop,
$g_0(A_\mu^a)_0=gA_\mu^a$. Therefore,}
\begin{align}
    g_0 = \mu^{\frac{4-n}{2}}\left(1+\frac{43}{6}\frac{g^2}{\ep}\right)g\,.
\end{align}
Finally, the renormalization relations for the quartic couplings and masses are:
\begin{align}
    \la_1^0 &= \mu^{4-n} \left[\la_1 - \frac{1}{\ep}\left({\frac{9}{8}} g^4 - 7\la_1 g^2 - \la_{12}\, g^2 + 20\la_1^2 + \la_{12}^2\right)\right]\,,\nn\\
    \la_2^0 &= \mu^{4-n} \left[\la_2 - \frac{1}{\ep}\left({\frac{9}{8}} g^4 - 7\la_2 g^2 - \la_{12}\, g^2 + 20\la_2^2 + \la_{12}^2\right)\right]\,,\nn\\
    (m_1^0)^2 &=m_1^2 + \frac{1}{\ep}\left[\frac{7}{2}m_1^2g^2 + m_2^2g^2 - 8m_1^2\la_1 - 2m_2^2\la_{12}\right]\,,\nn\\
    (m_2^0)^2 &= m_2^2 + \frac{1}{\ep}\left[\frac{7}{2}m_2^2g^2 + m_1^2g^2 - 8m_2^2\la_2 - 2m_1^2\la_{12}\right]\,.
\end{align}
Furthermore, the relation for the mixed coupling is given by
\begin{align}
    \la_{12}^0 = \mu^{4-n}\left[\la_{12} + \frac{1}{\ep}\left(-{\frac{9}{4}} g^4 + 5g^2\la_{12} + 4g^2(\la_1+\la_2)-8(\la_1+\la_2)\la_{12} - 4\la_{12}^2\right)\right]\,.
\end{align}
From these relations, we can find the beta functions and the anomalous dimension (gamma function) for the gauge coupling $g$. Since our focus is on evaluating the behavior of the quartic couplings, it will not be necessary to compute the beta functions for the masses. Thus,
\begin{align}\label{betag}
    \mu \frac{dg}{d\mu} = \frac{n-4}{2}\left[1 - \frac{43}{3}\frac{g^2}{\ep}\right]g \implies \be_g = \lim_{n\to4}\mu \frac{dg}{d\mu} = -\,\frac{43}{6(4\pi)^2}g^3
\end{align}
and the beta functions for the quartic and mixed couplings are given by
\begin{align}
    \be_{\la_1} &= \frac{1}{(4\pi)^2}\left[{\frac{9}{8}}g^4 +20\la_1^2 - 7g^2\la_1 + \la_{12}^2  - \la_{12}\,g^2\right]\, ,\nn\\
    \be_{\la_2} &= \frac{1}{(4\pi)^2}\left[{\frac{9}{8}}g^4 +20\la_2^2 - 7g^2\la_2 +  \la_{12}^2  - \la_{12}\,g^2\right]\,,\nn\\
    \be_{\la_{12}} &= \frac{1}{(4\pi)^2}\left[{\frac{9}{4}}g^4 + 4\la_{12}^2 - 5g^2\la_{12} - 4g^2(\la_1+\la_2) + 8(\la_1 + \la_2)\la_{12}\right]\,.
\end{align}
For the specific case $\la_1 = \la_2 = \frac12 \la_{12} = \la$, that is, when gauge invariance is preserved in the initial action, we have
\begin{align}
    \be_{\la_1} = \be_{\la_2} = \frac{1}{2} \be_{\la_{12}} = \frac{1}{(4\pi)^2}\left[24\la^2- 9g^2\la+{\frac{9}{8}}g^4 \right],
\end{align}
{
This expression agrees with the one-loop result of Cheng, Eichten, and Li~\cite{ChengEichtenLi1974} for the vector representation of $SU(N)$, presented in item 4 of Appendix B of that reference, after setting $N=2$ and accounting for the different normalization of the quartic interaction in the classical action, $\la_{\mathrm{CEL}}=2\la$. This agreement provides a consistency check of our calculation in the symmetric limit.}
In particular, for the gauge coupling constant $g$, we have
\begin{align}
    g^2(\mu) = g_0^2\left[{1+\frac{43g_0^2}{3(4\pi)^2}\ln\left(\frac{\mu}{\mu_0}\right)}\right]^{-1}\,,
\end{align}
where $g_0 = g(\mu_0)$. As expected, even with the explicit breaking of gauge symmetry in the scalar sector, Yang-Mills theories maintain their asymptotic freedom in the UV and their non-perturbative behavior in the IR. To solve the flow equations for the quartic couplings, we will employ the following change of variables \cite{ChengEichtenLi1974}:
\begin{align}\label{newvariables}
    \bar \la_1(t) = \frac{\la_1(t)}{g^2(t)} \,\,\text{ , }\,\,  \bar \la_2(t) = \frac{\la_2(t)}{g^2(t)}\,\,\text{ and }\,\, \bar{\la}_{12}(t) = \frac{\la_{12}(t)}{g^2(t)}\,.
\end{align}
Using these definitions, considering $t = \ln\left(\mu/\mu_0\right)$ and
\begin{align}
    \be_{\bar{\la}_i}g^2 + 2g\bar{\la}_i \be_g = \be_{\la_i} ,
\end{align}
we obtain
\begin{align}
  \be_{\bar{\la}_1} &= \frac{g^2}{(4\pi)^2} \left[{\frac{9}{8}} + 20\bar{\la}_1^2 + \frac{22}{3} \bar{\la}_1 +\bar\la_{12}^2 - \bar\la_{12}\,\right] \,, \nn\\
  \be_{\bar{\la}_2} &= \frac{g^2}{(4\pi)^2} \left[{\frac{9}{8}} + 20\bar{\la}_2^2 + \frac{22}{3} \bar{\la}_2  +\bar\la_{12}^2 - \bar\la_{12}\,\right]\,,\nn\\
   \be_{\bar\la_{12}} &= \frac{g^2}{(4\pi)^2}\left[{\frac{9}{4}} + 4\bar\la_{12}^2 + \frac{28}{3}\bar\la_{12} - 4(\bar\la_1+\bar\la_2) + 8(\bar\la_1 + \bar\la_2)\bar\la_{12}\right]\,.
\end{align}
We can rewrite the above equations in terms of a new parameter $\tau$,
\begin{align}\label{lambdatau}
    \frac{d\bar \la_1}{d\tau} &=  {\frac{9}{8}} + 20\bar{\la}_1^2 + \frac{22}{3} \bar{\la}_1 +\bar\la_{12}^2 - \bar\la_{12} \,\,, \nn\\
  \frac{d\bar \la_2}{d\tau} &=  {\frac{9}{8}} + 20\bar{\la}_2^2 + \frac{22}{3} \bar{\la}_2 +\bar\la_{12}^2 - \bar\la_{12} \,\,,\nn\\
  \frac{d\bar \la_{12}}{d\tau} &= {\frac{9}{4}} + 4\bar\la_{12}^2 + \frac{28}{3}\bar\la_{12} - 4(\bar\la_1+\bar\la_2) + 8(\bar\la_1 + \bar\la_2)\bar\la_{12}\,\,,
\end{align}
where
\begin{align}
    \tau = \frac{3}{43} \ln\left[1+\frac{43 g_0^2}{3(4\pi)^2}t\right]\,.
\end{align}
Note that in the UV limit, as $t \to +\infty$, the new variable behaves in the same manner, $\tau \to +\infty$. Let us define $\ze(\tau)$ as the difference between $ \bar \la_1(\tau)$ and $ \bar \la_2(\tau)$, that is, 
\begin{align}
    \ze(\tau) = \bar \la_1(\tau) - \bar \la_2(\tau)\,.
\end{align}
From Equations \eqref{lambdatau}, we find that 
\begin{align}\label{equationzeta}
    \frac{d\ze}{d\tau} = \ze \left(20\ze + 40 \bar \la_2 + \frac{22}{3}\right)\, \implies {\frac{d\log|\ze|}{d\tau}=20(\bar \la_1+\bar \la_2)+\frac{22}{3}}.
\end{align}

We focus on the region of positive quartic couplings, considering configurations compatible with the stability of the scalar potential. From Eq. \eqref{equationzeta}, using
$\zeta=\bar{\lambda}_1-\bar{\lambda}_2$, we obtain
\begin{align}\label{zetapositive}
    \frac{d\ln|\zeta|}{d\tau}
    =20(\bar{\lambda}_1+\bar{\lambda}_2)+\frac{22}{3}>0,
\end{align}
for $\zeta\neq0$ and $\bar{\lambda}_1,\bar{\lambda}_2\geq0$.

Therefore, the magnitude of the difference between the rescaled scalar self-couplings increases monotonically toward the UV and decreases toward the IR, as long as the RG flow remains within this region. This result shows that a small explicit symmetry breaking between the rescaled scalar self-couplings is amplified at higher energies and suppressed at lower energies.

However, this behavior does not necessarily imply a complete restoration of gauge symmetry in the deep infrared. Indeed, the one-loop running of the gauge coupling exhibits an infrared Landau pole, near which the perturbative analysis ceases to be reliable. Therefore, our conclusions are restricted to the region where the perturbative RG flow remains applicable.

\section{Fixed points}
From the beta functions for the quartic couplings, 
\begin{align}
    \frac{d\la_1}{dt} &= \frac{1}{(4\pi)^2}\left[{\frac{9}{8}}g^4 +20\la_1^2 - 7g^2\la_1 + \la_{12}^2 - \la_{12}\,g^2\right]\, ,\nn\\
    \frac{d\la_2}{dt} &= \frac{1}{(4\pi)^2}\left[{\frac{9}{8}}g^4 +20\la_2^2 - 7g^2\la_2 + \la_{12}^2- \la_{12}\,g^2\right]\,,\nn\\
     \frac{d\la_{12}}{dt} &= \frac{1}{(4\pi)^2}\left[{\frac{9}{4}}g^4 + 4\la_{12}^2 - 5g^2\la_{12} - 4g^2(\la_1+\la_2) + 8(\la_1 + \la_2)\la_{12}\right]\,,\nn\\
     \frac{d g}{dt} &= -\,\frac{43}{6(4\pi)^2}g^3\,,
\end{align}
the fixed points correspond to the values that satisfy the following relations:
\begin{align}\label{betalambda}
    & 20\la_1^2 + \la_{12}^2  = 0\,, \nn\\
    &20\la_2^2 + \la_{12}^2= 0\,,\nn\\
    &4\la_{12}^2+ 8(\la_1 + \la_2)\la_{12} = 0\,.
\end{align}
The only possible solution is the trivial fixed point
\[ (g,\la_1,\la_2,\la_{12}) \,=\,(0,0,0,0)\,.\]

However, to investigate the ultraviolet behavior of the scalar sector, we analyze the possible fixed points of the rescaled couplings defined in Eq. \eqref{newvariables}. From the beta functions for the rescaled couplings $\bar\la_i$, we obtain
\begin{align}\label{lambdatau1}
    &\frac{9}{8} + 20\bar{\la}_1^2
    + \frac{22}{3}\bar{\la}_1
    +\bar\la_{12}^2-\bar\la_{12}=0\,,
    \nonumber\\
    &\frac{9}{8} + 20\bar{\la}_2^2
    + \frac{22}{3}\bar{\la}_2
    +\bar\la_{12}^2-\bar\la_{12}=0\,,
    \nonumber\\
    &\frac{9}{4} + 4\bar\la_{12}^2
    + \frac{28}{3}\bar\la_{12}
    - 4(\bar\la_1+\bar\la_2)
    + 8(\bar\la_1+\bar\la_2)\bar\la_{12}=0\,.
\end{align}
The first equation can be rewritten as
\begin{align}
    20\left(\bar\la_1+\frac{11}{60}\right)^2
    +\left(\bar\la_{12}-\frac12\right)^2
    +\frac{73}{360}=0\,,
\end{align}
which shows that the system admits no real fixed points for any values of the rescaled quartic couplings.

Physically, this result reveals a fundamental aspect of the model, consistent with the behavior already observed in Eq. \eqref{zetapositive}: there are no finite UV fixed points for the rescaled quartic couplings. Furthermore, from Eq. \eqref{lambdatau} for $\bar\la_1$, considering $\bar\la_1\geq 0$, we obtain
\begin{align}
    \frac{d\bar\la_1}{d\tau}
    \geq \frac{7}{8}+20\bar\la_1^2\,,
\end{align}
which not only shows a monotonic increase of $\bar\la_1$ toward the UV, but also implies that the RG flow develops a divergence at a finite value of $\tau$.

Consequently, the scalar self-interactions prevent the complete asymptotic freedom of the theory, although the Yang-Mills gauge coupling remains asymptotically free within the one-loop approximation.

It is also worth noting the behavior of the model around the Gaussian (trivial) fixed point, where all scalar couplings vanish ($\lambda_1 = \lambda_2 = \lambda_{12} = 0$). Evaluating the exact beta functions from Eq. \eqref{betag} at this point, we observe that they do not vanish; instead, they become strictly positive due to the contributions from the vector sector:
\begin{equation}
    \beta_{\lambda_1}\Big|_{\lambda_i=0} = \beta_{\lambda_2}\Big|_{\lambda_i=0} = \frac{1}{(4\pi)^2} \left( {\frac{9}{8}}g^4 \right) > 0 \quad \text{and} \quad \beta_{\lambda_{12}}\Big|_{\lambda_i=0} = \frac{1}{(4\pi)^2} \left( {\frac{9}{4}}g^4 \right) > 0 \, .
\end{equation}

Physically, this implies that even if we assume a theory with no scalar self-interactions at the classical level, quantum fluctuations of the gauge bosons will generate them radiatively. Consequently, the Gaussian fixed point acts as a UV repulsor for the scalar sector. The gauge coupling acts as a driving force that pushes the scalar couplings away from the origin; as soon as the parameters $\lambda_i$ acquire non-zero values, their self-interactions (terms proportional to $\lambda^2$) take over the Renormalization Group flow, inevitably culminating in the finite-scale divergence (Landau pole) discussed earlier. This dynamics reaffirms the robustness of the derived equations, perfectly reproducing the asymptotic instability characteristic of scalar fields coupled to Yang-Mills theories.

\newpage
\section{Conclusions and discussions}
\label{sec6}

In this work, we investigated the one-loop quantum corrections and the renormalization group (RG) flow of the scalar couplings in a deformed $SU(2)$ Yang--Mills--Faddeev--Popov theory. The deformation was introduced by adding the action of a complex scalar doublet in the fundamental representation. We showed that this doublet can be mapped into a real $SO(4)$ multiplet, allowing us to split the scalar masses and couplings and explicitly break gauge symmetry.

The background field method, combined with the Schwinger--DeWitt technique \cite{DeWitt1965, DeWitt2003}, allowed us to compute the one-loop divergences of the deformed theory and demonstrate its renormalizability at this order. We simplified the calculations by setting the scalar and gauge background fields to zero at different stages. The recovariantization of the counterterms was then justified using Feynman-diagram calculations, the structure of the heat-kernel coefficients, and the mass-dimension analysis of the action.

While our previous investigation of this explicit symmetry-breaking mechanism in the Abelian sector (sQED) indicated a tendency toward asymptotic symmetry restoration in the infrared \cite{Lima2025}, we found a different behavior in the present non-Abelian model. Although the Yang--Mills gauge coupling remains asymptotically free \cite{Gross1973, Politzer1973}, the scalar self-interactions prevent this property from extending to the full set of couplings of the theory. The analysis of the one-loop beta functions, whose consistency was verified by recovering the known results in the gauge-invariant limit, showed that there are no real fixed points for the rescaled quartic couplings and that the RG flow develops a divergence at a finite energy scale. Furthermore, we found that the asymmetry between the rescaled scalar self-couplings increases toward the UV and decreases toward the IR, as long as the flow remains within the region of positive couplings.

These results allow us to distinguish two features of the RG evolution of the scalar couplings. The divergence of the flow in the UV occurs even in the symmetric regime and is therefore independent of explicit symmetry breaking. On the other hand, the difference between the rescaled scalar self-couplings increases monotonically toward higher energies, indicating an amplification of the asymmetry introduced in the scalar sector. This behavior follows directly from the RG equations, without assuming that the couplings are close to the symmetric regime. In the infrared, although this asymmetry is suppressed, we cannot conclude that gauge symmetry is completely restored, since the perturbative evolution is limited by the non-perturbative behavior of the gauge coupling.

\section*{Acknowledgements}
The author is grateful to Ilya L. Shapiro and Gustavo P. de Brito for their support and insightful discussions. This study was financed in part by the Coordenação de Aperfeiçoamento de Pessoal de Nível Superior - Brasil (CAPES) - Finance Code 001.

\section*{Appendix A}
\label{App}
In this section, we address the point mentioned in Section 5 concerning the recovariantization of the term $(\pa_\mu \phi)^\dagger(\pa_\mu\phi)$ into $(D_\mu \phi)^\dagger (D_\mu \phi)$. We focus on three points: the mass-dimension analysis of the counterterms, the structure of the heat-kernel coefficient $\hat{a}_2$, and the diagrammatic calculation of the three-point function $\langle A_\mu^a \phi^\dagger \phi \rangle$.

In the symmetric case, we know that the structure of the divergences must be covariant, i.e., we obtain a counterterm of the form $(D_\mu \phi)^\dagger (D_\mu \phi)$. Expanding this structure, we have
\begin{align}
(D_\mu\phi)^\dagger(D^\mu\phi)
&= (\pa_\mu\phi^\dagger)(\pa^\mu\phi)
-\frac{ig}{2}\left[(\pa_\mu\phi^\dagger)t^a A^{a\mu}\phi
-\phi^\dagger A_\mu^a t^a(\pa^\mu\phi)\right]
+\frac{g^2}{4}A_\mu^a A^{b\mu}\phi^\dagger t^a t^b\phi\,.
\end{align}
In our case with broken symmetry, deviations from this structure in the cubic and quartic terms should be proportional to $(m_1^2 - m_2^2)$, $(\la_1-\la_2)$ or $(2\la_{1,2} - \la_{12})$, or to mixed combinations, so that these deviations vanish in the symmetric limit and the covariant structure of the kinetic term is recovered. By canonical mass dimension, terms proportional to $(m_1^2 - m_2^2)$ are ruled out, since the corresponding cubic and quartic terms would have mass dimension six, requiring a coefficient of the form $1/m^2$. Such inverse powers of masses are excluded because the local UV divergences are polynomial in the masses.

It remains to show that these cubic and quartic terms cannot appear with factors of $\la_r$, $r = 1,2,12$, since these factors are dimensionless and cannot be ruled out by the mass dimension of the Lagrangian. For the cubic term, the only one-loop 1PI contribution involving a scalar quartic coupling is represented diagrammatically below.
\begin{figure}[H]
\begin{center}
\begin{tikzpicture}
\begin{feynman}

\vertex (a1) at (-1.5,0) {$A_\mu^a$};
\vertex [dot] (v1) at (0,0);
\vertex [dot] (v2) at (2,0);
\vertex (a2) at (3.5,1) {$\phi$};
\vertex (a3) at (3.5,-1) {$\phi^\dagger$};

\diagram* {
    (a1) -- [photon] (v1),
    (v1) -- [charged scalar, half left] (v2),
    (v2) -- [charged scalar, half left] (v1),
    (a2) -- [charged scalar] (v2),
    (v2) -- [charged scalar] (a3),
};

\end{feynman}
\end{tikzpicture}
\end{center}
\caption{Scalar-loop contribution to the
$A_\mu^a\phi^\dagger\phi$ vertex.}
\label{fig:Aphiphi-scalar-loop}
\end{figure}
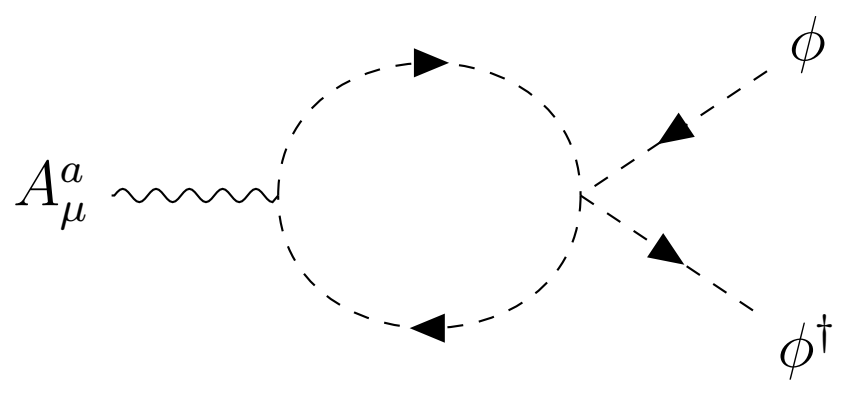
Up to overall factors, the contribution of this diagram is given by
\begin{align}
   I_\mu =  \la_r \,g \mu^{4-d}\int \frac{d^dp}{(2\pi)^d}\frac{2p_\mu + k_\mu}{(p^2+m_i^2)((p+k)^2+m_j^2)}, 
\end{align}
where the integral is written in Euclidean signature. Using the Feynman parametrization and the shift $p_\mu \to p_\mu - \al k_\mu$, we obtain
\begin{align}
    I_\mu = \la_r \,g \,\mu^{4-d}\int_0^1 d\al \int \frac{d^dp}{(2\pi)^d}\frac{2p_\mu+(1-2\al) k_\mu}{(p^2+\De^2)^2},
\end{align}
where $\Delta^2=k^2\al(1-\al)+(1-\al) m_i^2+ \al m_j^2$. By parity, the term in the numerator that is linear in $p_\mu$ vanishes. Thus, we only need to analyze the second contribution. To extract its UV divergence, we write
\begin{align}
     I_\mu = \la_r \,g\,\mu^{4-d}\int_0^1 d\al \,(1-2\al) \int \frac{d^dp}{(2\pi)^d}\left\{\frac{k_\mu}{(p^2+m_i^2)^2} -2\frac{k_\mu (\De^2-m_i^2)}{(p^2+m_i^2)^3}+\cdots\right\}.
\end{align}
Note that the integral over $\al$ vanishes for the first term, whereas the remaining terms are UV finite and contain $\al$-dependent integrals that do not necessarily vanish. Hence, this diagram does not contribute to the divergent part as $d\to4$ and does not generate UV-divergent contributions proportional to $\la_r$ in the cubic term.

Finally, we need to understand why the quartic term cannot be proportional to a scalar quartic coupling. For this purpose, let us note that the scalar sector of the bilinear form of the action can be written as
\begin{align}
    H_s = -D^2 + U_s
\end{align}
and the one-loop correction obtained through the Schwinger-DeWitt technique, up to total derivatives, is given by
\begin{align}
    \bar\Ga^{(1)}_{\text{div}} \supset -\frac{\mu^{n-4}}{\ep}\int d^nx\tr\left[\frac{1}{2}\hat{U}_s^2 + \frac{1}{12}\hat{S}_{\al\be}^2\right].
\end{align}
We only need to consider the scalar sector because, at one loop, the contributions to the mixed quartic vertex involving scalar quartic couplings contain only scalar loops. The operator $\hat{U}_s$ depends on quantities such as masses, quartic couplings and scalar background fields, but not on $A_\mu^a$, whereas $\hat{S}_{\mu\nu}$ is proportional to the background field strength $F_{\mu\nu}^a$ and the gauge coupling $g$. Thus, counterterms of the form $\la_r\,g^2A^2\phi^\dagger\phi$ do not arise, since no mixed product of these operators occurs in the trace. Therefore, the divergent correction to the mixed quartic vertex cannot be proportional to $\la_i$.

Consequently, in the divergent part at one loop, neither the cubic terms nor the quartic term can be proportional to $(m_1^2 - m_2^2)$, $(\la_1-\la_2)$ or $(2\la_{1,2} - \la_{12})$. Their coefficients therefore retain the same proportionality relations as in the symmetric theory, ensuring that divergences of the form $(\pa_\mu \phi^\dagger) (\pa_\mu \phi)$ can be mapped into $(D_\mu \phi)^\dagger (D_\mu \phi)$.

\section*{Appendix B}

In this section, we address the substitution $2\sqrt{\la_1\la_2}\to\la_{12}$, introduced to simplify the divergence calculations presented in Section 5.  As already discussed in the text, the perturbative Feynman rules of the unrestricted theory generate only nonnegative integer powers of the independent couplings. The square-root dependence in the restricted calculation originates from the relation $\la_{12}=2\sqrt{\la_1\la_2}$ imposed at the level of the action. Therefore, terms linearly proportional to $2\sqrt{\la_1\la_2}$ can be identified with contributions proportional to $\la_{12}$. 

For the contributions proportional to $4\la_1\la_2$ that arise in the quartic counterterms, the restriction makes terms proportional to $\la_{12}^2$ indistinguishable from those proportional to $\la_1\la_2$. To distinguish these possibilities, we examine the one-loop four-point diagrams. A contribution proportional to $\la_1\la_2$ would require one $(\phi_1^*\phi_1)^2$ vertex and one $(\phi_2^*\phi_2)^2$ vertex. Since the free scalar propagators do not connect the two species, these vertices cannot form a connected one-loop diagram.  For counterterms of the form $(\phi_i^*\phi_i)^2$, the scalar candy diagrams contain either two $(\phi_i^*\phi_i)^2$ vertices, generating contributions proportional to $\la_i^2$, or two mixed vertices, generating contributions proportional to $\la_{12}^2$. 

For counterterms of the form $(\phi_i^*\phi_i)(\phi_j^*\phi_j)$, with $i\neq j$, the scalar candy diagrams contain either two mixed vertices, generating contributions proportional to $\la_{12}^2$, or one mixed vertex and one $(\phi_i^*\phi_i)^2$ or $(\phi_j^*\phi_j)^2$ vertex, generating contributions proportional to $\la_i\la_{12}$ or $\la_j\la_{12}$, respectively. Thus, the terms proportional to $4\la_1\la_2$ in the restricted result must be identified with contributions proportional to $\la_{12}^2$.  

More generally, the one-loop quartic counterterm coefficients are quadratic polynomials in $\la_1$, $\la_2$, $\la_{12}$ and $g^2$. Any additional polynomial contribution that vanishes on the restricted surface $\la_{12}^2=4\la_1\la_2$ must therefore have the form $c(\la_{12}^2-4\la_1\la_2)$, where $c$ is independent of these couplings. Such a contribution would necessarily contain a term proportional to $\la_1\la_2$, which is excluded by the diagrammatic analysis above. Hence, $c=0$, eliminating this ambiguity in the reconstruction of the unrestricted one-loop result.

\newpage
\bibliographystyle{unsrt}
\bibliography{ref}
\end{document}